\documentclass[aps,pra,twocolumn,showpacs,superscriptaddress,longbibliography]{revtex4-2}
\usepackage{graphicx} 
\usepackage{amsmath}
\usepackage{graphicx,epstopdf}
\usepackage{gensymb}
\newcommand{\be}{\begin{equation}}
	\newcommand{\ee}{\end{equation}}
\newcommand{\bea}{\begin{eqnarray}}
	\newcommand{\eea}{\end{eqnarray}}
\newcommand{\bse}{\begin{subequations}}
	\newcommand{\ese}{\end{subequations}}

\usepackage{color}
\usepackage[colorlinks,bookmarks=false,citecolor=darkblue,linkcolor=red,urlcolor=blue]{hyperref}

\definecolor{darkred}{rgb}{0.7,0.0,0.0}

\definecolor{darkblue}{rgb}{0,0.02,0.45}

\definecolor{darkgreen}{rgb}{0.02,0.45,0.0}

\definecolor{violet}{rgb}{0.8,0.2,0.6}

\begin{document}

\title{Deformed spin-$\frac12$ square lattice in antiferromagnetic NaZnVOPO$_4$(HPO$_4$)}
\author{S. Guchhait}
\affiliation{School of Physics, Indian Institute of Science Education and Research Thiruvananthapuram-695551, India}
\author{D. V. Ambika}
\author{Qing-Ping Ding}
\affiliation{Ames Laboratory and Department of Physics and Astronomy, Iowa State University, Ames, Iowa 50011, USA}
\author{M. Uhlarz}
\affiliation{Dresden High Magnetic Field Laboratory (HLD-EMFL), Helmholtz-Zentrum Dresden-Rossendorf, 01328 Dresden, Germany}
\author{Y. Furukawa}
\affiliation{Ames Laboratory and Department of Physics and Astronomy, Iowa State University, Ames, Iowa 50011, USA}
\author{A. A. Tsirlin}
\email{altsirlin@gmail.com}
\affiliation{Felix Bloch Institute for Solid-State Physics, Leipzig University, 04103 Leipzig, Germany}
\author{R. Nath}
\email{rnath@iisertvm.ac.in}
\affiliation{School of Physics, Indian Institute of Science Education and Research Thiruvananthapuram-695551, India}
\date{\today}

\begin{abstract}
We report the structural and magnetic properties of a new spin-$\frac12$ antiferromagnet NaZnVOPO$_4$(HPO$_4$) studied via x-ray diffraction, magnetic susceptibility, high-field magnetization, specific heat, and $^{31}$P nuclear magnetic resonance (NMR) measurements, as well as density-functional band-structure calculations. While thermodynamic properties of this compound are well described by the $J_1-J_2$ square-lattice model, \textit{ab initio} calculations suggest a significant deformation of the spin lattice. From fits to the magnetic susceptibility we determine the averaged nearest-neighbor and second-neighbor exchange couplings of $\bar J_1\simeq -1.3$\,K and $\bar J_2\simeq 5.6$\,K, respectively. Experimental saturation field of 15.3\,T is consistent with these estimates if 20\% spatial anisotropy in $J_1$ is taken into account. Specific heat data signal the onset of a magnetic long-range order at $T_{\rm N} \simeq 2.1$~K, which is further supported by a sharp peak in the NMR spin-lattice relaxation rate. The NMR spectra mark the superposition of two P lines due to two noneqivalent P sites where the broad line with the strong hyperfine coupling and short $T_1$ is identified as the P(1) site located within the magnetic planes, while the narrow line with the weak hyperfine coupling and long $T_1$ is designated as the P(2) site located between the planes.
\end{abstract}

\maketitle
\section{\textbf{Introduction}}
Low-dimensional spin systems augmented with strong frustration reveal suppression of a conventional magnetic order and may lead to a quantum disordered ground state like quantum spin liquid (QSL)~\cite{Savary016502,*Balents199}. Two-dimensional (2D) spin-$1/2$ frustrated square-lattice (FSL) is a well-known example where frustration appears because of the competition between nearest-neighbor (NN, $J_1$) and next-nearest-neighbor (NNN, $J_2$) exchange interactions along the edges and diagonals of a square, respectively ($J_1-J_2$ model). Theoretical studies have determined a global phase diagram with different ground states depending on the sign and relative strength of the exchange couplings ($\alpha=\frac{J_2}{J_1}$)~\cite{Shannon599,Schmidt214443}. Ferromagnetic (FM), N\'{e}el antiferromagnetic (NAF), and degenerate columnar antiferromagnetic (CAF) ordered phases are stabilized and extended in the regimes of $-0.5<\alpha \le -\infty$, $-\infty\le\alpha <0.5$, and $0.5<\alpha <-0.5$, respectively.
Different quantum disordered phases with novel order parameters are predicted to emerge at the phase boundaries or the quantum critical regimes.
For instance, QSL~\cite{Zhang067201,Hu060402,Wang107202}, plaquette valence-bond solid (PVBS)~\cite{Gong027201,Doretto104415} or columnar valence-bond solid (CVBS)~\cite{Haghshenas174408} states are expected around $\alpha\simeq 0.5$ while a spin nematic phase is predicted for $\alpha\simeq -0.5$ in the phase diagram~\cite{Shannon027213}.

In the recent past, a handful number of Cu$^{2+}$ (3$d^{9}$), V$^{4+}$ (3$d^{1}$), and Mo$^{5+}$ (4$d^{1}$) based spin-$1/2$ FSL magnets have been studied as $J_1-J_2$ model compounds. The vanadates are the most celebrated ones and include Li$_2$VO$X$O$_4$ ($X$= Si, Ge) with $J_1, J_2>0$ and $J_2\gg J_1$~\cite{Melzi1318,Rosner2002}, along with V$^{4+}$ phosphates that typically show $J_1<0$ and $J_2>0$, also resulting in the CAF ground state~\cite{Nath214430,Roy012048,Tsirlin174424,Bossoni014412,Tsirlin014429,Nath064422}. A few other vanadates, such as Zn$_2$VO(PO$_4$)$_2$~\cite{Yogi024413}, VOMoO$_4$~\cite{Kiani075112,Bombardi220406,Carretta094420}, and PbVO$_3$~\cite{Tsirlin092402}, feature $J_1\gg J_2$ and lie in the NAF region of the phase diagram. A further variability becomes possible with Cu$^{2+}$ as well as $4d$ Mo$^{5+}$ compounds that span both $J_2\gg J_1$ and $J_2\ll J_1$ limits~\cite{Ishikawa064408,Takeda104406,Xu105801,Vasala496001,Koga054426,Babkevich237203,Walker064411,Watanabe2022contrasting,Guchhait57006}.
Unfortunately, none of these compounds fall in the quantum critical regimes of the phase diagram around $\alpha=\frac12$. Moreover, some of them show intricate deformations of the magnetic square lattice, because the underlying crystal symmetry is lower than tetragonal~\cite{Tsirlin214417}. This deformation was verified experimentally in several V$^{4+}$ phosphates, including Pb$_2$VO(PO$_4)_2$~\cite{Bettler184437}, SrZnVO(PO$_4)_2$~\cite{Landolt224435}, and BaCdVO(PO$_4)_2$~\cite{Bhartiya144402}. Each of them showed an interesting pre-saturation phase~\cite{Povarov024413,Bhartiya033078,Skoulatos014405,Landolt094414,Landolt224435}, which is reminiscent of the nematic phase of the $J_1-J_2$ model~\cite{Shannon027213}, but deviations from the ideal square lattice put into question the applicability of this theoretical scenario, and alternative interpretations were indeed proposed recently~\cite{Ranjith2021nmr,Landolt2022}.


\begin{figure*}
	\includegraphics[width=\linewidth]{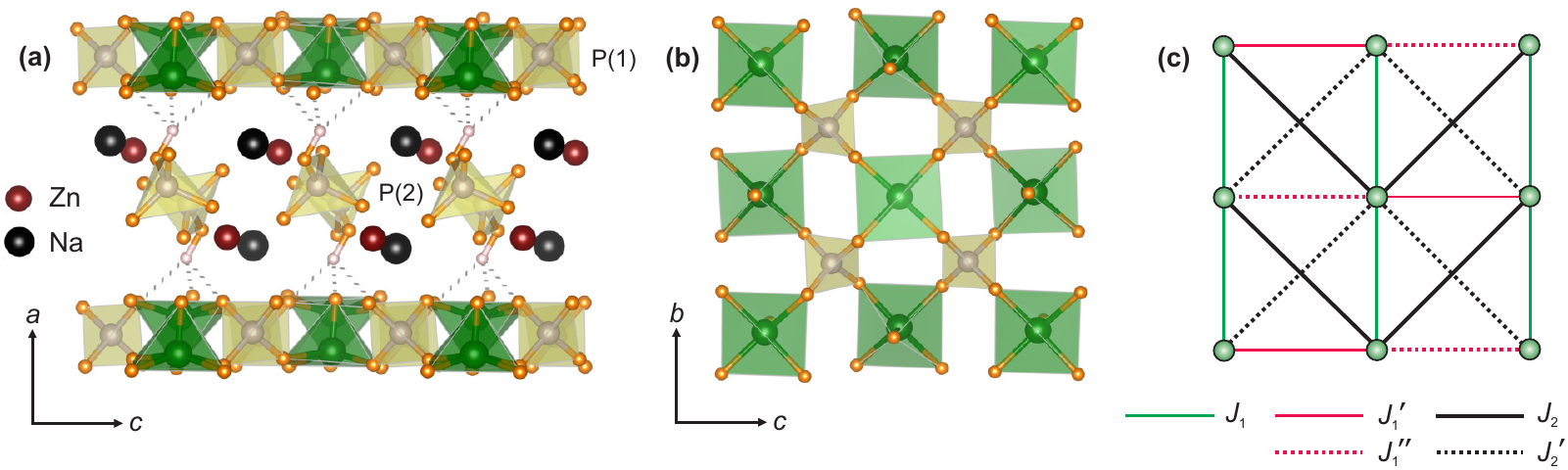} 
	\caption{(a) Crystal structure of NaZnVOPO$_4$(HPO$_4$) with the hydrogen position and hydrogen bonds (dashed lines) determined in this work. (b) Magnetic [VOPO$_4$] layer with the square-lattice-like arrangement of the V$^{4+}$ ions. (c) Deformed square lattice with 5 nonequivalent interactions. \texttt{VESTA} software~\cite{vesta} was used for crystal structure visualization.}
	\label{Fig1}
\end{figure*}

In this context, finding V$^{4+}$ square-lattice compounds with different magnitudes of the deformation is important. Here, we report low-temperature magnetic behavior of the hitherto unexplored NaZnVOPO$_4$(HPO$_4$) as a \mbox{spin-$\frac12$} square-lattice candidate with a different spacer separating the magnetic layers and, potentially, a different regime of exchange couplings compared to the widely studied $AA'$VO(PO$_4)_2$ phosphates ($AA'$ = Pb$_2$, SrZn, BaCd). 
The monoclinic crystal structure of NaZnVOPO$_4$(HPO$_4$) features V$^{4+}$ ions in the square-pyramidal coordination. They are joined into layers via P(1)O$_4$ tetrahedra (Fig.~\ref{Fig1}), while the interlayer space is filled with the Na$^+$ and Zn$^{2+}$ ions, as well as the P(2)O$_4$ tetrahedra that represent the HPO$_4$ groups.

\section{Methods}
Pale-blue coloured powder of the titled compound was synthesized by the conventional hydrothermal method. 0.150~g Na$_2$CO$_3$ (Aldrich, 99.995\%), 0.154~g V$_2$O$_5$ (Aldrich, 99.99\%), and 0.115~g Zn (Aldrich, 99\%) powders were mixed with 5~ml of a 1.5~M aqueous solution of H$_3$PO$_4$, sealed in a 23~ml Teflon lined bomb, and heated at 240$^{o}$C for 8 days followed by slow cooling (10$\degree$C/hour) to room temperature. The obtained blue color product was washed carefully with distilled water and dried in an oven maintained at $100\degree$C for 24~hours. To check the phase purity of the compound, powder x-ray diffraction (XRD) experiment was performed using a PANalytical powder diffractometer with Cu~$K_{\alpha}$ radiation ($\lambda_{\rm avg} \simeq 1.5418$~\AA) at room temperature. We have also performed temperature-dependent powder XRD measurements on the pure phase powder sample in the temperature range 15~K$\leq T\leq 300$~K, using a low-temperature attachment (Oxford PheniX) to the x-ray diffractometer. Le-Bail fit of the powder XRD patterns was performed using the FULLPROF software package~\cite{Carvajal55}, taking the initial structural parameters from the previous report~\cite{LeFur1735}. 

Temperature variation of magnetization ($M$) was measured in the temperature range 1.8–350 K in different magnetic fields ($H$) using a SQUID magnetometer (MPMS3, Quantum Design). The isothermal magnetization was measured at $T = 1.8$~K from 0 to 7~T. The high-field magnetization measurement was performed at $T=1.4$~K in pulsed magnetic fields up to 30~T using the facility at the Dresden High Magnetic Field Laboratory~\cite{Skourski214420,Tsirlin132407}.
The temperature dependent specific heat of this sample was measured on a sintered pellet in a large temperature range (0.5~K$\leq T \leq 200$~K) using the Physical Property Measurement System (PPMS, Quantum Design) and adopting the thermal relaxation technique. For measurements below 2~K, $^{3}$He attachment to the PPMS was used.

The Nuclear Magnetic Resonance (NMR) measurements were carried out on the $^{31}$P nuclei (gyromagnetic ratio $\frac{\gamma_{\rm N}}{2\pi}=17.237$~MHz/T and nuclear spin $I=\frac{1}{2}$) in the temperature range 1.8~K$\leq T\leq 250$~K. The NMR spectra at different temperatures were obtained by varying the magnetic field at a constant frequency of 121~MHz. The spin-lattice relaxation rate ($1/T_1$) was measured by the single saturation pulse method at two frequencies (30.2~MHz and 121~MHz). The temperature-dependent NMR shift, $K(T)= [H_{\rm ref}/H(T)-1]$, was calculated from the resonance field of the sample $H$ with respect to the resonance field of a non-magnetic reference sample $(H_ {\rm ref})$.

Magnetic couplings were determined by density-functional-theory (DFT) band-structure calculations performed in the \texttt{FPLO} code~\cite{fplo} using experimental structural parameters from Ref.~\cite{LeFur1735}, except the hydrogen position that was optimized as further explained in Sec.~\ref{sec:dft}. The Perdew-Burke-Ernzerhof (PBE) approximation for the exchange-correlation potential was employed~\cite{pbe96}. We used superexchange theory in the vein of Kugel-Khomskii model~\cite{Mazurenko2006,Tsirlin2011b}, as well as the mapping approach~\cite{Xiang2011,Tsirlin2014} based on total energies of collinear spin configurations obtained from DFT+$U$ calculations with the on-site Coulomb repulsion $U_d=4$\,eV, Hund's coupling $J_d=1$\,eV, and double-counting correction in the atomic limit~\cite{Weickert2016}. 

Experimental thermodynamic properties were modeled by high-temperature series expansion (HTSE) for the $J_1-J_2$ spin-$\frac12$ square lattice~\cite{Rosner014416} as well as full diagonalization (FD) for the $4\times 4$ finite lattice with periodic boundary conditions. 
Additionally, quantum Monte Carlo (QMC) simulations were performed for the nonfrustrated \mbox{spin-$\frac12$} square lattice using directed loop algorithm~\cite{Alet036706} in the stochastic series expansion (SSE)~\cite{SandvikR14157} representation. QMC simulations were performed for $16\times 16$ finite lattices with periodic boundary conditions, using $4\times10^{4}$ sweeps and $4\times10^{3}$ thermalization sweeps. The \texttt{ALPS} package~\cite{BauerP05001} was used for both FD and QMC. 

\section{Results}
\subsection{X-ray Diffraction}
\begin{figure}[h]
\includegraphics[width=\linewidth]{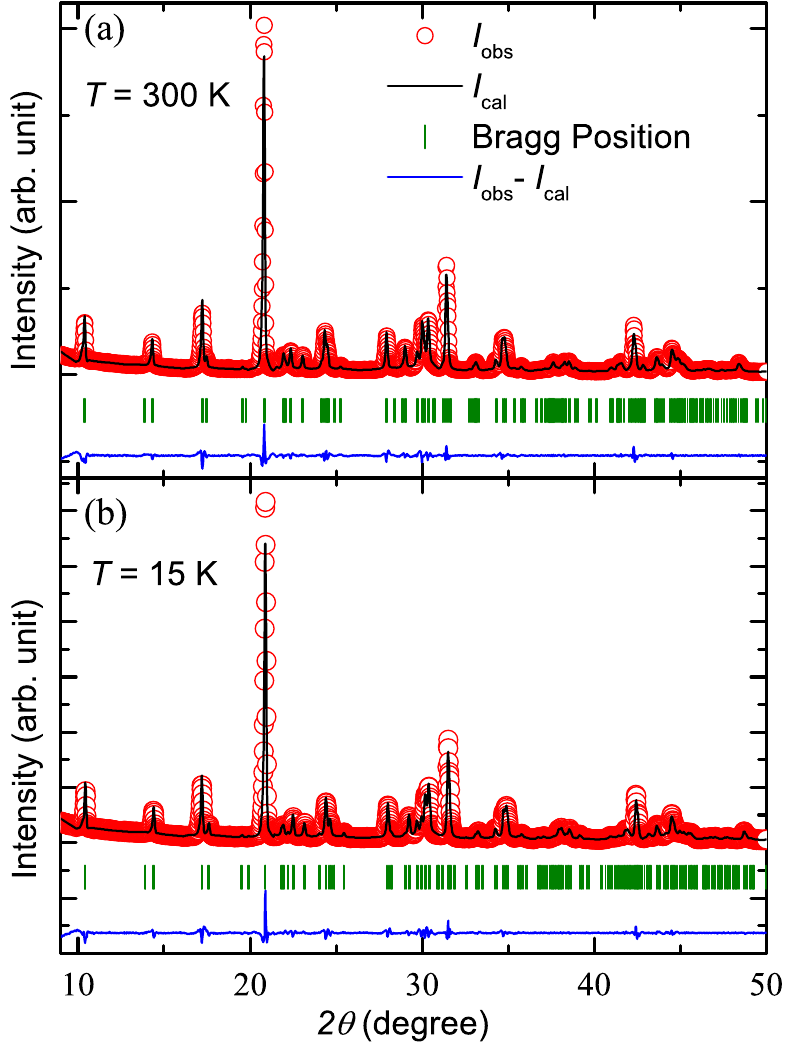}
\caption{Powder XRD patterns (open circles) at (a) $T=300$~K and (b) $T=15$~K. The solid line is the Le-Bail fit, the vertical bars mark the expected Bragg peak positions, and the lower solid line corresponds to the difference between the observed and calculated intensities. The goodness-of-fit is achieved to be $\chi^{2}\sim5.4$ and $\sim6.2$ for $T=300$~K and $15$~K, respectively.}
\label{Fig2}
\end{figure}
\begin{figure}[h]
	\includegraphics[width=\linewidth]{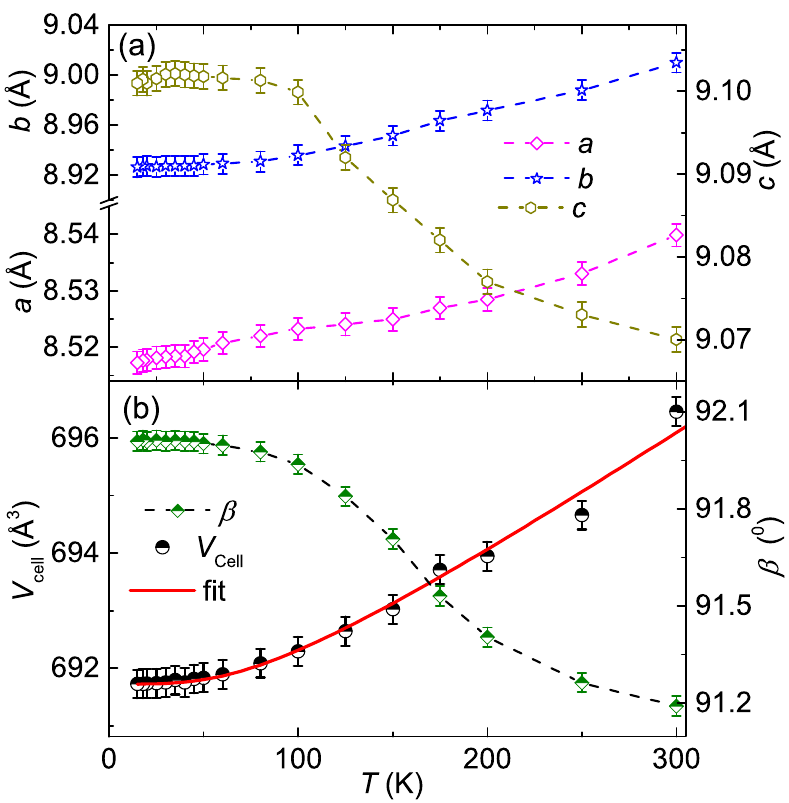}
	\caption{(a) Lattice constants ($a$, $b$, and $c$) as a function of temperature. (b) Monoclinic angle ($\beta$) along with the unit cell volume ($V_{\rm cell}$) are plotted as a function of temperature from 15~K to 300~K. The solid line represents the fit of $V_{\rm cell}$ using Eq.~\eqref{Vcell}.}
	\label{Fig3}
\end{figure}
The powder XRD patterns of NaZnVOPO$_4$(HPO$_4$) are analyzed by Le Bail fits. Figure~\ref{Fig2} presents the data at two end temperatures ($T=300$~K and 15~K). The entire XRD pattern down to 15~K could be indexed using the monoclinic crystal structure with the space group $P2_1/c$. Neither any indication of structural transition nor lattice distortion is found down to 15~K. The refined lattice parameters and unit cell volume ($V_{\rm cell}$) are [$a=8.5418(4)$~\AA, $b=8.9937(5)$~\AA, $c=9.0765(5)$~\AA, $\beta=91.186(3)^\circ$, and $V_{\rm cell}\simeq696.46$~\AA$^3$] and [$a=8.5190(2)$~\AA, $b=8.9281(4)$~\AA, $c=9.1029(4)$~\AA, $\beta=92.013(5)^\circ$, and $V_{\rm cell}\simeq691.73$~\AA$^3$] for $T=300$~K and 15~K, respectively. The obtained lattice parameters at room temperature are in close agreement with the values reported earlier~\cite{LeFur1735}. 
The temperature dependence of the lattice parameters ($a$, $b$, $c$, and $\beta$) and $V_{\rm cell}$ are presented in Fig.~\ref{Fig3}. The lattice constants $a$ and $b$ are found to decrease in a systematic way with decreasing temperature while $c$ and monoclinic angle $\beta$ increase with decreasing temperature and then rearch a plateau. These lead to an overall thermal contraction of the $V_{\rm cell}$ with temperature.

The variation of $V_{\rm cell}$ with temperature can be expressed in terms of the internal energy [$U(T)$] of the system~\cite{Wallace1998,Budd18,*Guchhait224415},
\begin{equation}\label{Vcell}
	V_{\rm cell}(T) = \frac{\gamma U(T)}{K_0} + V_0.
\end{equation}
Here, $V_0$ is the unit-cell volume at $T= 0$~K, $\gamma$ is the Gr\"uneisen parameter, and $K_0$ is the bulk modulus of the system. According to the Debye model, $U(T)$ can be written as,
\begin{equation}\label{Uenergy}
		U(T) = 9Nk_{\rm B}T\left(\frac{T}{\theta_{\rm D}}\right)^3 \int_{0}^{\frac{\theta_{\rm D}}{T}}\frac{x^3}{(e^{x}-1)}dx,
\end{equation}
where $N$ is the total number of atoms per unit cell, $k_{\rm B}$ is the Boltzmann constant, and 
$\theta_{\rm D}$ is the Debye temperature~\cite{Kittel2004}. The variable $x$ inside the integration stands for the quantity $\frac{\hbar\omega}{k_{\rm B}T}$ with phonon frequency $\omega$ and Planck constant $\hbar$. Here, $\theta_{\rm D}=\frac{\hbar\omega_{\rm D}}{k_{\rm B}}$ and $\omega_{\rm D}$ is the upper limit of $\omega$.
 The best fit of the $V_{\rm cell}(T)$ data using Eq.~\eqref{Vcell} [solid line in Fig.~\ref{Fig3}(b)] yields the parameters: $\theta_{\rm D}\simeq312$~K, $V_0\simeq$ 691.72~\AA$^3$, and $\frac{\gamma}{K_0}\simeq8.84\times10^{-12}$~Pa$^{-1}$.

\subsection{Magnetization} 
\begin{figure}
	\includegraphics[width=\linewidth]{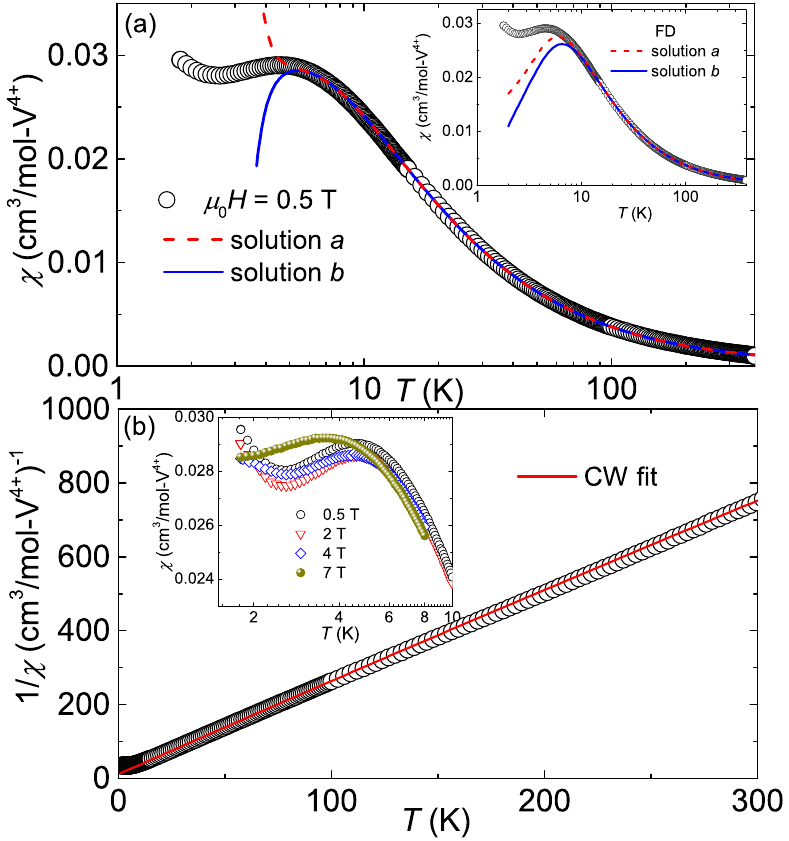}
	\caption{(a) $\chi(T)$ measured in the magnetic field of $H = 0.5$~T. The lines (solid and dashed) are the HTSE fits using the isotropic spin-$\frac12$ FSL model, with two different solutions (solution-$a$ and solution-$b$). Inset: The simulated $\chi(T)$ using the FD method taking the $J_1$ and $J_2$ values from solutions-$a$ and $b$, respectively. (b) $1/\chi$ vs $T$. The red solid line is the CW fit. Inset: $\chi(T)$ in the low-$T$ region measured in different fields.}
	\label{Fig4}
\end{figure}
The magnetic susceptibility [$\chi(T)\equiv M/H$] of NaZnVOPO$_4$(HPO$_4$) measured in an applied field of $H=0.5$~T is shown in Fig.~\ref{Fig4}(a). In the high-temperature region, $\chi(T)$ follows a typical Curie-Weiss (CW) behaviour and shows a rounded maximum at $T_{ \chi}^{\rm max}\simeq4.6$~K. Such a maximum represents the short-range AFM order in the low-dimensional spin systems. Below about 2.6~K, a small upturn is likely due to paramagnetic impurities and/or defects present in the powder sample. No trace of magnetic long-range-order (LRO) is detected down to 2~K. We have also measured $\chi(T)$ in different applied fields [inset of Fig.~\ref{Fig4}(b)] but no obvious features associated with magnetic LRO is found, except the suppression of a broad maximum towards low temperatures.

The inverse susceptibility, $1/\chi(T)$, is shown in Fig.~\ref{Fig4}(b). In the paramagnetic regime, $1/\chi(T)$ was fitted by the CW law
\begin{equation}\label{cw}
\chi(T) = \chi_0 + \frac{C}{T - \theta_{\rm CW}},
\end{equation}
where the first term ($\chi_0$) represents the combination of temperature-independent diamagnetic and Van-Vleck paramagnetic susceptibilities. The second term is the CW law where $C$ is the Curie constant and $\theta_{\rm CW}$ is the CW temperature. The fit above 25~K returns the parameters: $\chi_0\simeq4.26\times10^{-5}$~cm$^3$/mol-V$^{4+}$, $C\simeq0.393$~cm$^3$K/mol-V$^{4+}$, and $\theta_{\rm CW}\simeq -4.3$~K. Using the value of $C$, the effective moment is calculated to be $\mu_{\rm eff}$ $[=(3k_{\rm B}C/N_{\rm A}\mu_{\rm B}^{2})^{\frac{1}{2}}$, where $N_{\rm A}$ is the Avogadro’s number and $\mu_{\rm B}$ is the Bohr magneton] $\simeq1.77~\mu_{\rm B}$/V$^{4+}$. This value of $\mu_{\rm eff}$ is close to the actual value $1.73~\mu_{\rm B}$ for a \mbox{spin-$\frac12$} transition-metal ion with $g=2$. The negative value of $\theta_{\rm CW}$ indicates the dominant AFM exchange coupling between the V$^{4+}$ ions.
The core diamagnetic susceptibility ($\chi_{\rm core}$) of the compound caused by the core orbital electrons was calculated to be $-1.37\times10^{-4}$~cm$^3$/mol by adding the $\chi_{\rm core}$ of Na$^{+}$, Zn$^{2+}$, V$^{4+}$, P$^{5+}$, and O$^{2-}$ ions~\cite{selwood2013}. The Van-Vleck paramagnetic susceptibility ($\chi_{\rm vv}$) was calculated to be $\sim1.8 \times10^{-4}$~cm$^3$/mol by subtracting $\chi_{\rm core}$ from $\chi_0$, which is very close to the value reported for other V$^{4+}$ based compounds~\cite{Mukharjee224403}.

As evident from the structural data (Fig.~\ref{Fig1}), the system deviates from the isotropic square lattice. This spatial anisotropy is mainly due to five non-equivalent exchange couplings, three between nearest neighbors ($J_1$, $J_1^{\prime}$, and $J_1^{\prime\prime}$) and two between next-nearest neighbors ($J_2$ and $J_2^{\prime}$). Previous studies suggest that this anisotropy has only a minor effect on thermodynamic properties at higher temperatures~\cite{Tsirlin214417}. Therefore, fits with the isotropic $J_1-J_2$ model return averaged values of the NN and NNN couplings, $\bar{J}_1=(2J_1 + J_1^{\prime} + J_1^{\prime\prime})/4$ and $\bar{J}_2=(J_2 + J_2^{\prime})/2$, respectively. We fitted the data with 
\begin{equation}
	\label{chi_twoD}
	\chi(T) = \chi_0 + \chi_{\rm spin}(T)
\end{equation}
using the temperature-independent term ($\chi_0$) and the 9th order HTSE [$\chi_{\rm spin}(T)$],
\begin{eqnarray}
	\label{chi_spin}
	\chi_{\rm spin}(T)=\frac{N_{\rm A}g^{2}\mu_{\rm B}^{2}}{k_{\rm B}T}\sum_{n}\left(\frac{\bar J_{1}}{k_{\rm B}T}\right)^{n}\sum_{m}c_{mn}\left(\frac{\bar J_{\rm 2}}{\bar J_{\rm 1}}\right)^{m}.
\end{eqnarray}
The values of the coefficients $c_{mn}$ are taken from Ref.~\cite{Rosner014416}. Our fit for $T \geq 7$~K yields two solutions: solution-$a$ with $\chi_0 \simeq 6 \times 10^{-5}$~cm$^{3}$/mol-V$^{4+}$, $g \simeq 2.04$, $\bar{J}_1 \simeq 5.18$~K, $\bar{J}_2 \simeq -0.66$~K and solution-$b$ with $\chi_0 \simeq 6.6 \times 10^{-5}$~cm$^{3}$/mol-V$^{4+}$, $g \simeq 2.04$, $\bar{J}_1 \simeq -1.28$~K, $\bar{J}_2 \simeq 5.59$~K.

\begin{figure}
	\includegraphics[width=\linewidth]{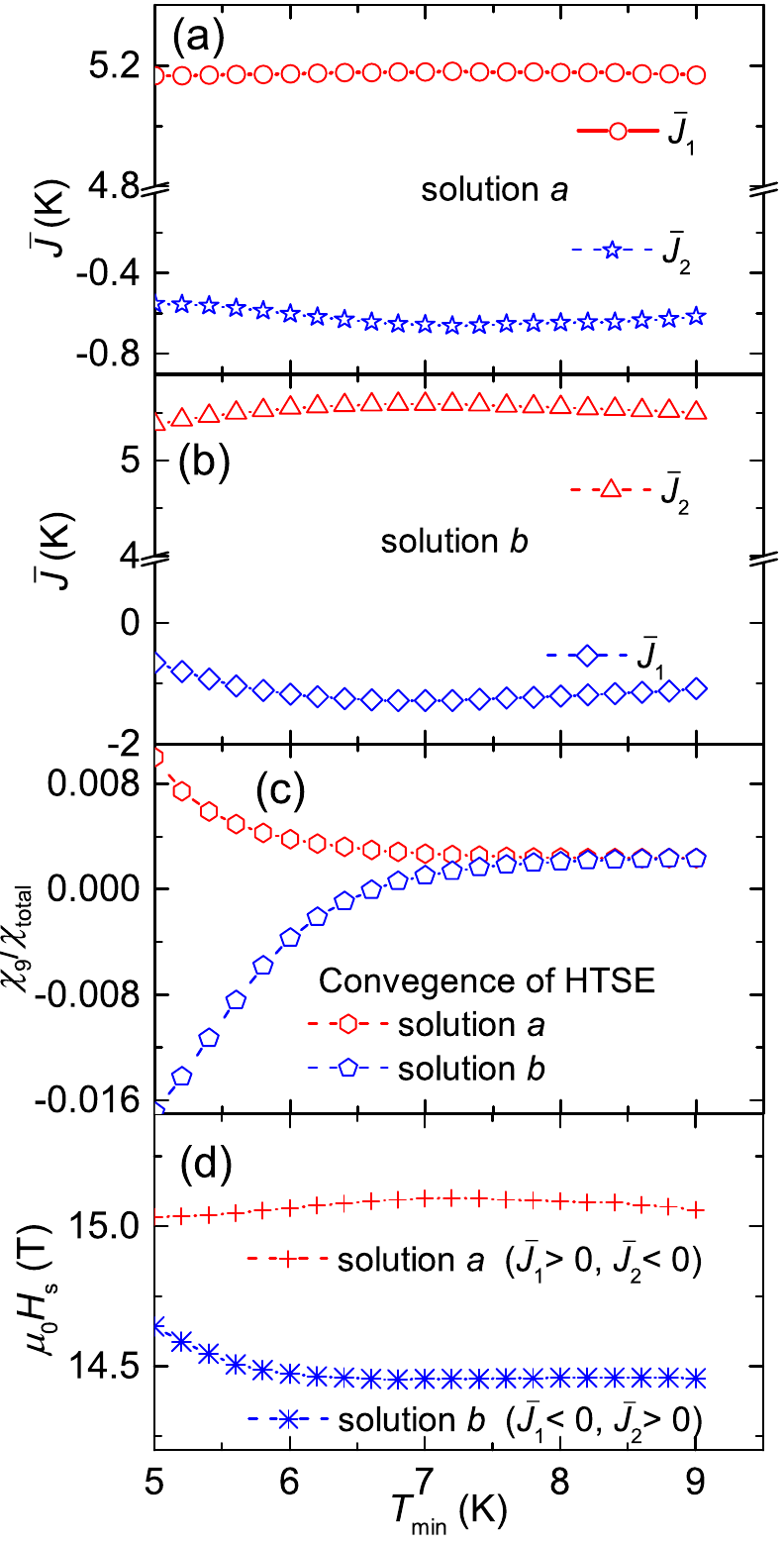}
	\caption{Results of the $\chi(T)$ fit using HTSE for isotropic spin-$\frac12$ FSL model by varying the minimum temperature of the fitting range ($T_{\rm min}$). (a) and (b) Averaged exchange interactions ($\bar{J}_1$ and $\bar{J}_2$) vs $T_{\rm min}$ for the solutions $a$ and $b$, respectively. (c) Convergence test of the HTSE fit for both solutions, $\chi_{9}(T)/\chi_{\rm total}(T)$ vs $T_{\rm min}$. (d) Saturation fields for the solutions $a$ and $b$ vs $T_{\rm min}$.} 
	\label{Fig5}
\end{figure}

The main difficulty of the HTSE fit is choosing the appropriate $T$-range. The convergence of the HTSE depends on the $\bar J_2/\bar J_1$ ratio, hence the lower limit of the fitting range ($T_{\rm min}$) should be chosen with caution depending on the results of the fit~\cite{Tsirlin174424}. 
For a precise estimation of the $J$ values, we varied the lower limit of the fitting $T$-range ($T_{\rm min}$) between 5 and 9~K and estimated the exchange couplings for both the solutions. Figures~\ref{Fig5}(a) and (b) present the variation of $\bar J_1$ and $\bar J_2$ with $T_{\rm min}$ for the solutions-$a$ and $b$, respectively. To check the convergence of the HTSE, we calculated the ratio of the ninth-order term to the total susceptibility [$\chi_{9}(T)/\chi_{\rm total}(T)$] using the parameters obtained from HTSE fits with a different $T_{\rm min}$. Both solutions are stable above 7\,K. We also performed the FD simulations of $\chi(T)$ using the values of $\chi_{0}$, $g$, $\bar J_1$, and $\bar J_2$ and found a good agreement with the experimental data [inset of Fig.~\ref{Fig4}(a)].

\begin{figure}
	\includegraphics[width=\linewidth]{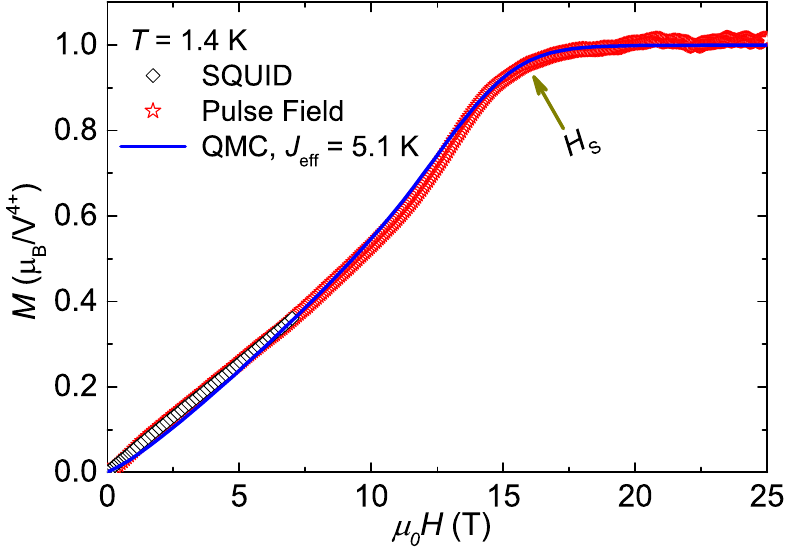}
	\caption {Magnetization ($M$) vs field ($H$) at $T=1.4$~K measured using pulse magnetic field and scaled with respect to the SQUID data. The solid line represents the QMC simulation, assuming the spatially isotropic nonfrustrated spin-$1/2$ square-lattice model with $J_{\rm eff}=5.1$~K. The arrow indicates the position of the saturation field ($H_{\rm S}$).} 
	\label{Fig6}
\end{figure}
Discriminating between the solutions-$a$ and $b$ may be possible using measurements of the saturation field~\cite{Tsirlin132407}. The $M$ vs $H$ data measured up to 25~T in pulsed magnetic fields at $T=1.4$~K are shown in Fig.~\ref{Fig6}. We have scaled the high-field data with respect to the magnetic isotherm at $T = 1.8$~K measured up to 7~T using a SQUID magnetometer. $M$ increases linearly with $H$ in the low-field region, shows a positive curvature in the intermediate fields, and then saturates at around $H_{\rm S} \simeq 15.3$~T. The positive curvature is typical for low-dimensional and frustrated spin systems~\cite{Thalmeier104441}.

Saturation field of an FSL magnet depends on the type of magnetic order. In the NAF case (solution-$a$), $H_{\rm S}=4\bar{J}_1k_{\rm B}/(g\mu _{\rm B})\simeq 15.1$\,T in good agreement with the experiment. On the other hand, in the CAF case $H_{\rm S}=(\mathcal{J}_1+2\bar{J}_2)2k_{\rm B}/(g\mu_{\rm B})$ where $\mathcal J_1$ is the weaker of the couplings $J_1$ and $(J_1'+J_1'')/2$. In the CAF state, spins align antiferromagnetically along the direction of this coupling and ferromagnetically along the orthogonal direction. Therefore, only the weaker coupling enters the saturation field. Using $\mathcal J_1=\bar J_1$ leads to $H_{\rm S}^{\rm CAF} \simeq 14.45$\,T for the solution-$b$, lower than in the experiment. On the other hand, the actual value of $\mathcal J_1$ is reduced owing to the deformation of the square lattice. Assuming 20\% spatial anisotropy in the NN couplings ($\mathcal J_1=0.8\bar J_1$) according to the DFT results (Sec.~\ref{sec:dft}), we arrive at $H_{\rm S}^{\rm CAF}\simeq 15.15$\,T, which is on par with the result for the solution-$a$ and also matches the experimental value.


Interestingly, the magnetization curve can be also well described by the simple NN square-lattice model with an effective coupling $J_{\rm eff}=5.1$\,K (Fig.~\ref{Fig6}), which is in good agreement with the leading exchange couplings extracted from the HTSE fits. 


\subsection{Specific Heat}
\begin{figure}
	\includegraphics[width=\linewidth]{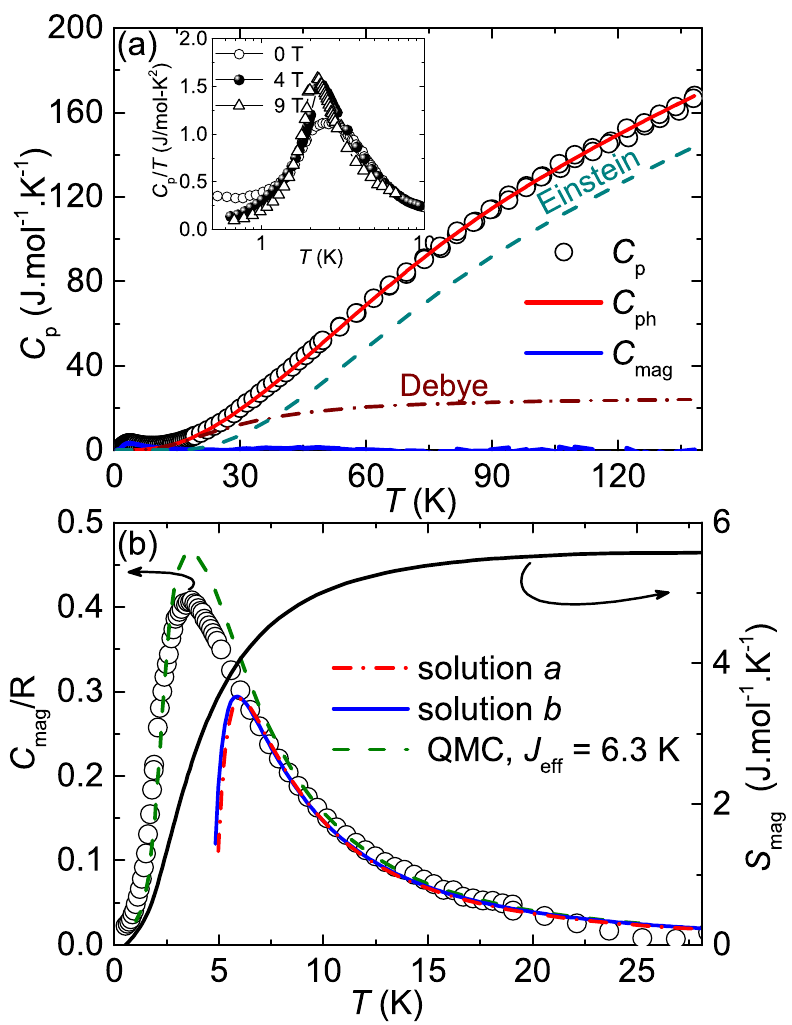}
	\caption{(a) Variation of $C_{\rm p}$ with temperature in the absence of magnetic field. The red solid line represents the simulated phonon contribution ($C_{\rm ph}$) taking into account the Debye (dash-dotted line) and Einstein (dashed line) terms. Inset: $C_{\rm p}/T$ vs $T$ at low temperatures and in different applied fields. The blue solid line is the magnetic contribution ($C_{\rm mag}$). (b) $C_{\rm mag}/R$ vs $T$ (left $y$-axis). The solid line represents the change in the magnetic entropy, $S_{\rm mag}$ vs $T$ (right $y$-axis). The red (dash-dotted) and blue (solid) lines are the HTSE fits corresponding to the two solutions, respectively. The green (dashed) line is the QMC simulation for an isotropic non-frustrated spin-$1/2$ square lattice model with $J_{\rm eff} = 6.3$~K.}
	\label{Fig7}
\end{figure}
In Fig.~\ref{Fig7}(a), we have plotted the temperature-dependent specific heat [$C_{\rm p}(T)$] measured from 0.5 to 140~K in zero applied field. It decreases systematically with temperature and passes through a broad maximum at $T_{C}^{\rm max} \simeq 3.7$~K, typical for low-dimensional oxides. An anomalous behavior was observed in a narrow temperature range $T_{\rm N} = 2 - 2.2$~K where we couldn't stabilize the temperature. This is a possible indication of the onset of a magnetic LRO.

In a magnetic insulator, the total specific heat $C_{\rm p}(T)$ is the sum of two main contributions: one is the phonon/lattice contribution [$C_{\rm ph}(T)$], which dominates in the high-temperature region, and another one is the magnetic contribution [$C_{\rm mag}(T)$], which dominates in the low-temperature region depending upon the strength of the exchange coupling.
In order to bring out the magnetic part of the specific heat, we first quantified the lattice contribution and then subtracted it from the total specific heat. We simulated the high-temperature $C_{\rm ph}(T)$ data taking into account the sum of one Debye [$C_{\rm D}(T)$] and three Einstein [$C_{\rm E}(T)$] terms, i.e., $C_{\rm ph}(T) = C_{\rm D}(T) + \sum\limits_{i=1}^{3} C_{\rm Ei}(T)$. The Debye and Einstein terms are expressed as
 \begin{equation}\label{Debye}
 	C_{\rm D}(T) = 9n_{\rm D}R\left(\frac{T}{\theta_{\rm D}}\right)^3 \int_{0}^{\frac{\theta_{\rm D}}{T}}\frac{x^4e^x}{(e^{x}-1)^2}dx
 \end{equation}
and
\begin{equation}\label{Einstein}
	C_{\rm Ei}(T) = 3n_{\rm Ei}R \left(\frac{\theta_{Ei}}{T}\right)^2\frac{e^{\left(\frac{\theta_{Ei}}{T}\right)}}{\left[e^{\left(\frac{\theta_{Ei}}{T}\right)}-1 \right]^2}.	
\end{equation}
Here, the Einstein temperatures $\theta_{\rm Ei}= \frac{\hbar\omega_{Ei}}{k_{\rm B}}$, $\omega_{Ei}$ are the respective Einstein frequencies, and $R$ denotes the universal gas constant. The values of $n_{\rm D}$ and $n_{\rm Ei}$ are chosen in such a way that the sum $n_{\rm D}+\sum\limits_{i=1}^{3}n_{\rm Ei}$ matches with the total number of atoms per formula unit. The best fit of the $C_{\rm p}$ data in the high-temperature regime using one Debye and three Einstein branches yields the characteristic temperatures: $\theta_{\rm D}\simeq 122$~K, $\theta_{\rm E1}\simeq307$~K, $\theta_{\rm E2}\simeq850$~K, and $\theta_{\rm E3}\simeq161$~K with $n_{\rm D}=1$, $n_{\rm E1}=5$, $n_{\rm E2}=7$, and $n_{\rm E3}=2$, respectively~\cite{Sebastian064413}. The red solid line in Fig.~\ref{Fig7}(a) is the total phononic contribution to the specific heat ($C_{\rm ph}$) extrapolated down to low temperature, while the dashed and dash-dotted lines are the Einstein and Debye contributions, respectively. The value of $\theta_{\rm D}$ and the average value of $\theta_{\rm E}$ match well with $\theta_{\rm D}$ estimated from the $V_{\rm cell}$ vs $T$ analysis.

$C_{\rm mag}$ estimated after subtracting $C_{\rm ph}$ from $C_{\rm p}$ is plotted in Fig.~\ref{Fig7}(b) as a function of temperature. The pronounced broad maximum at $T_C^{\rm max}\simeq 3.7$~K mimics short-range antiferromagnetic correlations. The change in magnetic entropy [$S_{\rm mag}(T)$] is obtained by integrating $C_{\rm mag}/T$ over temperature. It saturates to a value of $\sim 5.6$~J.mol$^{-1}$.K$^{-1}$ at around $T\simeq 25$~K, which is close to the expected value $R\ln2=5.76$~J.mol$^{-1}$.K$^{-1}$ for a two-level (spin-$\frac{1}{2}$) system, thus justifying our subtraction procedure and the evaluation of $C_{\rm mag}$.

The magnetic LRO is highlighted by plotting $C_{\rm p}/T$ vs $T$ in the inset of Fig.~\ref{Fig7}(a) in different applied fields. While the anomaly associated with the magnetic LRO is not pronounced in the zero-field data, the peak at $T_{\rm N} \simeq 2.1$~K becomes more pronounced with increasing the field. This is due to the transfer of entropy from the broad maximum to the transition anomaly. Surprisingly, no visible shift in $T_{\rm N}$ is perceived even with the field change of 9~T, indication of a robust AFM transition.

One can estimate the exchange couplings by analyzing $C_{\rm mag}(T)$ using the HTSE of a spin-$\frac12$ FSL model~\cite{Rosner014416},
\begin{eqnarray}
	\label{Cmag_spin}
	\frac{C_{\rm mag}(T)}{R}=\frac{\bar J_1}{k_{\rm B}T}\sum_{n}(-n)\left(\frac{\bar J_{1}}{k_{\rm B}T}\right)^{n}\sum_{m}e_{mn}\left(\frac{\bar J_{\rm 2}}{\bar J_{\rm 1}}\right)^{m}
\end{eqnarray}
We have fitted the $C_{\rm mag}(T)/R$ data at $T > 6$~K [see Fig.~\ref{Fig7}(b)] and arrived at two solutions that strongly resemble the two solutions from the susceptibility fit: solution-$a$ ($\bar{J}_1/k_{\rm B} \simeq 5.9$~K, $\bar{J}_2/k_{\rm B} \simeq -0.2$~K) and solution-$b$ ($\bar{J}_1/k_{\rm B} \simeq -1.6$~K, $\bar{J}_2/k_{\rm B} \simeq 6.1$~K). Alternatively, we can compare our experimental data with the QMC simulation for the nonfrustrated square-lattice model [see Fig.~\ref{Fig7}(b)]. The position of the maximum in $C_{\rm mag}$ is reproduced with $J_{\rm eff}=6.3$\,K, which is notably higher than in the $M(H)$ fit (Fig.~\ref{Fig6}). Moreover, the maximum value of $C_{\rm mag}(T)/R$ is higher than in the experiment. The simple NN square-lattice model is thus insufficient to describe the magnetic behavior of NaZnVOPO$_4$(HPO$_4$). Both frustration and deformation of the square lattice affect thermodynamic properties of this compound.

\subsection{$^{31}$ P NMR}
NMR is a convenient local probe of both static and dynamic properties. In NaZnVOPO$_4$(HPO$_4$), the P(1) site is strongly coupled to the V$^{4+}$ ions within the VOP(1)O$_4$ layer, while the P(2) site which is located between the adjacent layers is weakly coupled to the V$^{4+}$ ions [see Fig.~\ref{Fig1}(a)]. This difference allows a useful comparison and highlights the magnetic behavior of the square-lattice planes.

\subsubsection{$^{31}$P NMR Spectra}
\begin{figure}[h]
	\includegraphics[width=\linewidth]{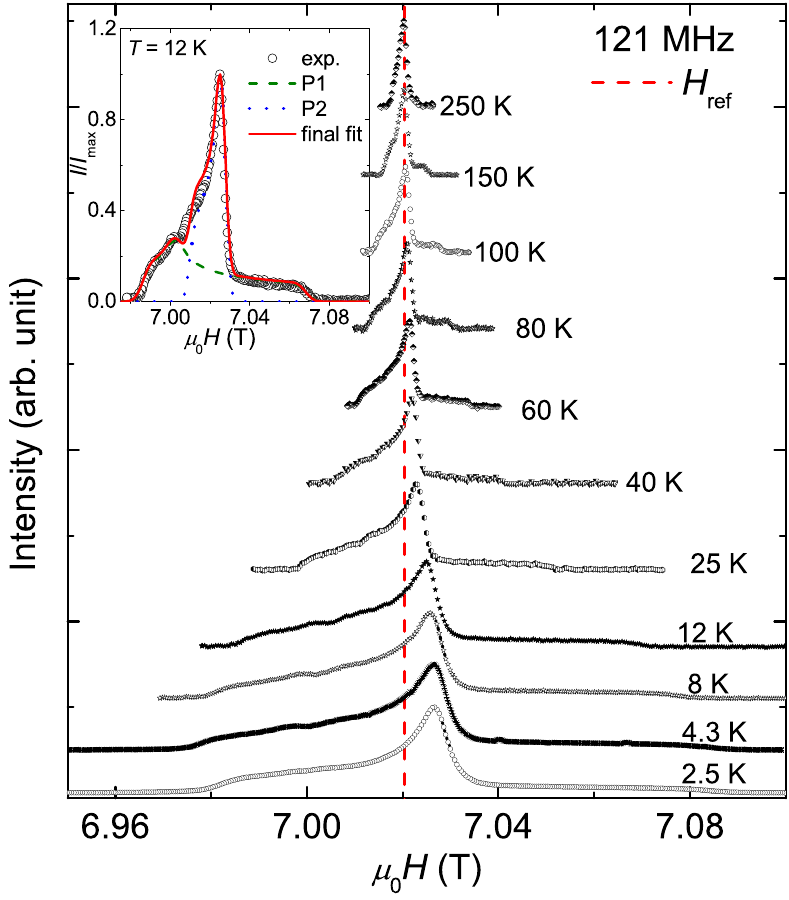}
	\caption {Temperature evolution of the $^{31}$P NMR spectra of NaZnVOPO$_4$(HPO$_4$) measured at 121~MHz. The dashed line indicates the reference field position. Inset: $^{31}$P NMR spectrum at $T = 12$~K with the dashed and dotted lines are the fits of the P(1) and P(2) sites, respectively, and the solid line (final fit) is the superposition of the P(1) and P(2) fits. The NMR shift values along the $x$, $y$, and $z$ directions, obtained from the fitting are $K_{x}^{(1)}\simeq0.48\%, K_{y}^{(1)}\simeq-0.68\%,$ and $K_{z}^{(1)}\simeq0.24\%$ for P(1) site and $K_{x}^{(2)}\simeq0.13\%, K_{y}^{(2)}\simeq-0.1\%$, and $K_{z}^{(2)}\simeq-0.09\%$ for P(2) site, respectively.}
	\label{Fig8}
\end{figure}
The field-sweep $^{31}$P NMR spectra above $T_{\rm N}$ measured in a radio frequency of 121~MHz are shown in Fig.~\ref{Fig8}. Each spectrum is normalized by its maximum amplitude and offset vertically by adding a constant. At high temperatures, the line is found to be narrow but asymmetric and the central peak appears at the zero-shift position. As the temperature is lowered, the line width increases drastically and becomes more anisotropic with two shoulders on either side of the central peak. This abnormal spectral shape can be attributed to two nonequivalent P sites in the crystal structure~\cite{LeFur1735}. Remarkably, the complete shape of the spectra could be reproduced considering the superposition of two spectral lines. The inset of Fig.~\ref{Fig8} portrays the spectral fit at $T=12$~K. We have a narrow central line with weak anisotropy and a broad asymmetric background with two distinct shoulders. 

The narrow central peak shifts weakly, whereas the shoulders move significantly with decreasing temperature. Thus, the narrow central line with a weak shift can be assigned to the P(2) site, which is weakly coupled while the broad line with the strong temperature-dependent behavior corresponds to the in-plane P(1) site, which is strongly coupled to the V$^{4+}$ spins. The asymmetric shape of both P(1) and P(2) sites is likely due to the anisotropy in $\chi(T)$ or asymmetry in the hyperfine coupling constant between the P nuclear spins and the V$^{4+}$ electronic spins. The overall spectral shape matches exactly with the spectral shape reported for other V$^{4+}$-based FSL compounds, $AA'$VO(PO$_4$)$_2$ ($AA'$ = Pb$_2$, SrZn), on the polycrystalline sample~\cite{Nath214430,Bossoni014412}.

\begin{figure}[h]
	\includegraphics[width=\linewidth]{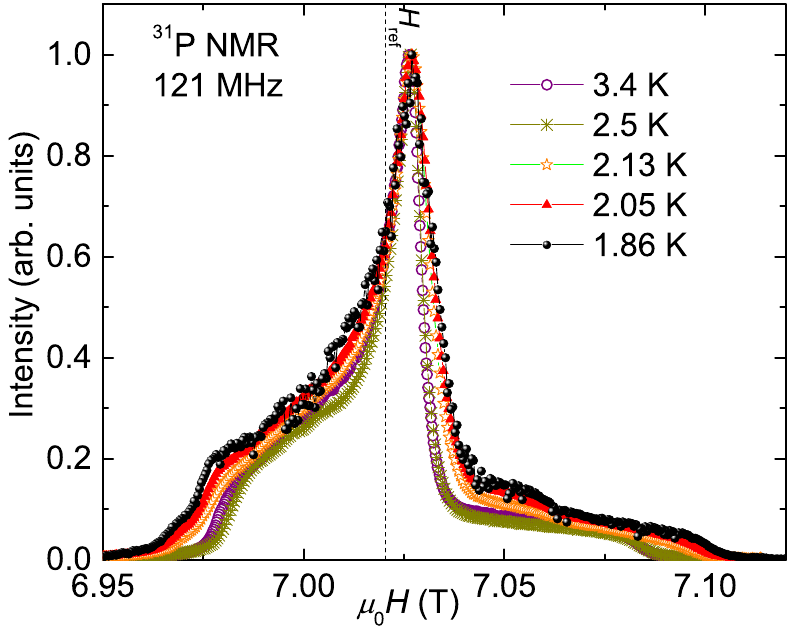}
	\caption {Temperature-dependent $^{31}$P NMR spectra of NaZnVOPO$_4$(HPO$_4$) measured at 121~MHz around $T_{\rm N}$. The dashed line indicates the reference field position.}
	\label{Fig9}
\end{figure}
The nature of the magnetic ordering can also be gleaned from the analysis of NMR spectra below $T_{\rm N}$. Figure~\ref{Fig9} shows the normalized $^{31}$P NMR spectra measured at 121~MHz around $T_{\rm N}$. Neither significant change in the line shape nor any visible line broadening are observed below $T_{\rm N}$. This indicates that the $^{31}$P site experiences only a weak static field in the ordered state. This observation is quite opposite to that reported for Pb$_2$VO(PO$_4$)$_2$~\cite{Nath214430} but similar to the $^{29}$Si NMR results on Li$_2$VOSiO$_4$ where local field on the Si site cancels out~\cite{Melzi2001}. All these compounds show a similar type of magnetic layers and a similar mutual arrangement of the V$^{4+}$ ions and PO$_4$/SiO$_4$ tetrahedra. However, only in Pb$_2$VO(PO$_4)_2$ the magnetic [VOPO$_4$] layers are strongly buckled~\cite{Tsirlin214417}. This may explain why the hyperfine couplings to V$^{4+}$ spins with the opposite alignment do not lead to the cancellation of the local field. On the other hand, in NaZnVOPO$_4$(HPO$_4$) the layers are almost flat (Fig.~\ref{Fig1}), and the absence of the $^{31}$P NMR line broadening below $T_{\rm N}$ can be still ascribed to the filtering of the hyperfine fields due to the AFM spin alignments in the ground state.


\subsubsection{$^{31}$P NMR Shift}
\begin{figure}[h]
	\includegraphics[width=\linewidth]{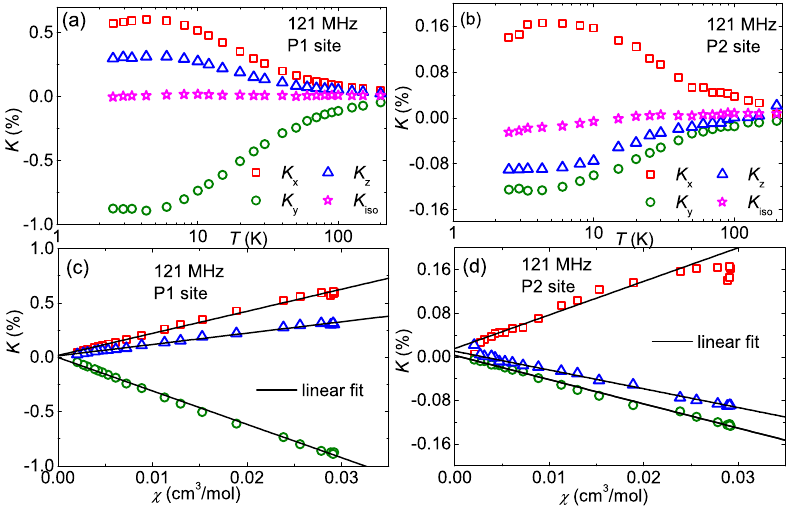}
	\caption {Anisotropic components of $K$ along the $x$, $y$, and $z$-directions as a function of temperature for (a) P(1) and (b) P(2) sites, respectively. $K$ vs $\chi$ measured at 7~T are plotted for all the three orientations for (c) P(1) and (d) P(2) sites, respectively. The solid lines are the straight line fits.}
	\label{Fig10}
\end{figure}
The spectrum at each temperature was fitted following the same procedure as described above for $T=12$~K. The estimated temperature-dependent NMR shift [$K(T)$] along different orientations ($K_{\rm x}$, $K_{\rm y}$, and $K_{\rm z}$) for both P(1) and P(2) sites are plotted in Fig.~\ref{Fig10}(a) and (b), respectively. All the components of $K(T)$ exhibit a broad maximum/minimum at around 4.3~K, an indication of the 2D AFM short-range-ordering. Further, the magnitude of $K(T)$ for the P(2) site is weaker than the P(1) site, as expected. The isotropic NMR shift was calculated as $K_{\rm iso} = (K_x+K_y+K_z)/3$, which is found to be almost temperature-independent for both the $^{31}$P sites. This also suggests that the isotropic part of hyperfine coupling at the P site from the four neighboring V$^{4+}$ spins is nearly averaged out.

As $K(T)$ is an intrinsic measure of the spin susceptibility $\chi_{\rm spin}(T)$ and is free from extrinsic contributions, one can write
\begin{equation}
	\label{K}
	K(T) = K_0 +\frac{A_{\rm hf }}{N_{\rm A}}\chi_{\rm spin}(T).
\end{equation}
Here, $K_0$ is the temperature-independent chemical (orbital) shift and $A_{\rm hf}$ is the hyperfine coupling constant between the $^{31}$P nucleus and V$^{4+}$ spins. In order to estimate $A_{\rm hf}$, we have plotted $K$ vs $\chi$, assuming temperature as an implied variable in Fig.~\ref{Fig10}(c) and (d) for P(1) and P(2) sites, respectively. In every case, $K$ vs $\chi$ plot is linear in the whole temperature range. A straight line fit results $A^{(1)}_{\rm hf} = (1134 \pm 19)$~Oe/$\mu_{\rm B}$, $(-1716 \pm 20)$~Oe/$\mu_{\rm B}$, and $(582.59 \pm 11)$~Oe/$\mu_{\rm B}$ along the $x$, $y$ and $z$-directions for the P(1) site and $A^{(2)}_{\rm hf}=(343 \pm 13)$~Oe/$\mu_{\rm B}$, $(-248 \pm 4)$~Oe/$\mu_{\rm B}$, and $(-192 \pm 3)$~Oe/$\mu_{\rm B}$ along the $x$, $y$ and $z$-directions for the P(2) site, respectively. Clearly, the magnitude of $A^{(1)}_{\rm hf}$ is almost one order of magnitude larger than $A^{(2)}_{\rm hf}$ in all the three directions proving the stronger coupling for P(1) than P(2).

For the P1 site, the transferred hyperfine coupling mainly arises from the interactions with the four nearest-neighbor V$^{4+}$ spins in the plane. The
isotropic and anisotropic transferred hyperfine couplings
originate from P(3s)-O(2p)-V(3d) and P(3p)-O(2p)-V(3d)
covalent bonds, respectively. Since P1 is surrounded by four
V$^{4+}$ ions forming a nearly square lattice in the plane,
the experimentally observed asymmetry in hyperfine
field indicates inequivalent P(3p)-O(2p)-V(3d)
bonds for the four NN V$^{4+}$ ions and hence a distortion in the square lattice, consistent with the low symmetry of the crystal structure as pointed out earlier.

\subsubsection{Spin-lattice relaxation rate $^{31}1/T_1$}
\begin{figure}[h]
	\includegraphics[width=\linewidth]{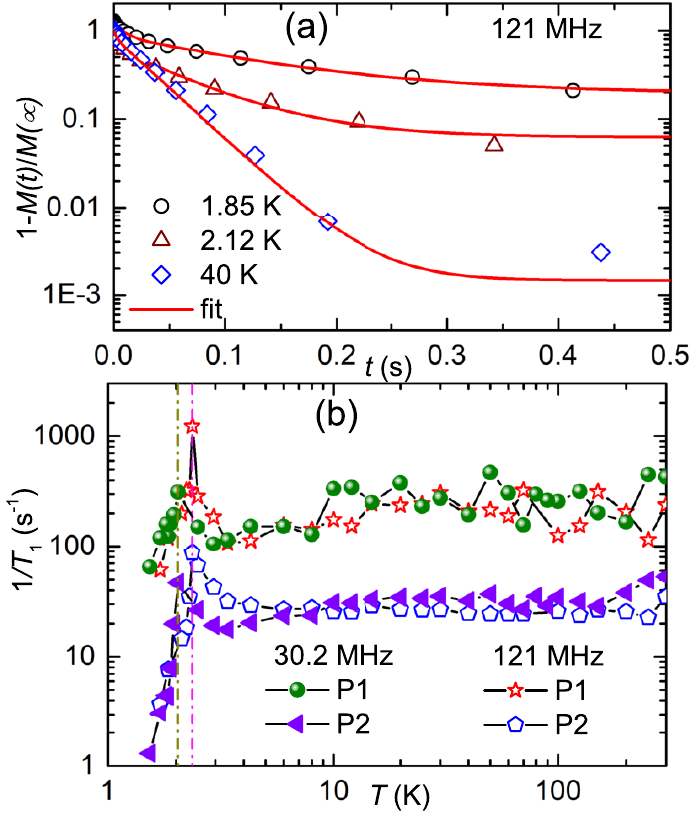}
	\caption{(a) Longitudinal recovery curves at three selective temperatures and the solid lines are fits using Eq.~\eqref{exp}. (b) $^{31}$P spin-lattice relaxation rates for P(1) ($1/T_{11}$) and P(2) ($1/T_{12}$) sites as a function of temperature measured in 30.2~MHz and 121~MHz. The data are shown in log-log scale in order to highlight the peak at $T_{\rm N}$. The vertical dash-dotted lines indicate the magnetic LRO at $T_{\rm N}\simeq 2.04$~K and $T_{\rm N}\simeq 2.2$~K for 30.2~MHz and 121~MHz data, respectively.}
	\label{Fig11}
\end{figure}
We have performed $^{31}$P spin-lattice relaxation rate ($1/T_1$) measurements as a function of temperature down to 1.7~K in two different frequencies, 30.2~MHz and 121~MHz. The recovery of the longitudinal nuclear magnetization after a saturation pulse could be fitted by a double exponential function
\begin{equation}
1-\frac{M(t)}{M(\infty)}= Ae^{-t/T_{11}}+Be^{-t/T_{12}},
\label{exp}
\end{equation}
where $M(t)$ is the nuclear magnetization at a time $t$ after the saturation pulse and $M(\infty)$ is the equilibrium nuclear magnetization. As each NMR spectrum is a superposition of two P-sites, which are inseparable, a double exponential function is used to fit the recovery curves. In Eq.~\eqref{exp}, $1/T_{11}$ and $1/T_{12}$ are the spin-lattice relaxation rates for P(1) and P(2) sites, respectively, and $A$ and $B$ account for their respective weight factors. At 121~MHz, the total spectral width was large and we were not able to saturate the whole spectrum using a single saturation pulse. Therefore, $1/T_1$ is also measured at a lower frequency of 30.2~MHz where the spectral width is reduced significantly and we were able to saturate the whole spectrum above $T_{\rm N}$.

The extracted $1/T_{11}$ and $1/T_{12}$ as a function of temperature at 30.2~MHz and 121~MHz and for both the P-sites are presented in Fig.~\ref{Fig11}(b). In the high-temperature ($T\geq4.3$~K) region, $1/T_1$ for both the P-sites are almost constant, typically expected in the paramagnetic regime~\cite{Moriya23}. At low temperatures, $1/T_{11}$ and $1/T_{12}$ show sharp peaks at around $T_{\rm N} \simeq 2.04$~K and 2.36~K for 30.2~MHz and 121~MHz, respectively, indicating the slowing down of fluctuating moments as we approach the magnetic LRO. Below $T_{\rm N}$, both $1/T_{11}$ and $1/T_{12}$ decrease toward zero due to the scattering of magnons by the nuclear spins~\cite{Islam174432}.

\subsection{Microscopic magnetic model}
\label{sec:dft}
Before discussing the magnetic model, we determine the position of hydrogen, which may be crucial for the correct evaluation of exchange couplings~\cite{Lebernegg2013}. In hydrophosphates, one expects a deformation of PO$_4$ tetrahedra because one of the oxygens is linked to hydrogen and should thus weaken its bond to phosphorous in order to keep the overall bond valence unchanged. Such a deformation is observed in the P(2)O$_4$ tetrahedra with the P--O bond distances of 1.490 [O(1)], 1.530 [O(7)], 1.537 [O(2)], and 1.585\,\r A [O(9)]~\cite{LeFur1735}. We thus considered different hydrogen positions in the vicinity of O(9) and found the lowest energy for hydrogen located at (0.7658,0.2823,0.7588) and separated from oxygen by 0.988\,\r A. The O--H bond is directed toward the [VOPO$_4$] layer, with the H atom forming three hydrogen bonds of $2.2-2.4$\,\r A to oxygen atoms of the VO$_5$ pyramids and P(1)O$_4$ tetrahedra (Fig.~\ref{Fig1}).

\begin{table}
\caption{\label{tab:exchange}
Interatomic V--V distances (in \r A) and exchange couplings (in K). The FM and AFM contributions are obtained from the superexchange model, Eq.~\eqref{eq:kk}, whereas total exchange couplings $J_i$ are calculated via the DFT+$U$ mapping procedure and thus deviate from $J_i^{\rm FM}+J_i^{\rm AFM}$. 
}
\begin{ruledtabular}
\begin{tabular}{cc@{\hspace{2em}}cc@{\hspace{2em}}r}
& $d_{\rm V-V}$ & $J_i^{\rm AFM}$ & $J_i^{\rm FM}$ & $J_i$ \\
 $J_1$   & 4.616 & 1.4  & $-5.6$ & $-5.1$ \\
 $J_1'$  & 4.542 & 1.4  & $-7.4$ & $-7.9$ \\
 $J_1''$ & 4.785 & 2.6  & $-3.1$ & $-1.0$ \\
 $J_2$   & 6.380 & 18.7 & $-0.4$ & 8.1 \\
 $J_2'$  & 6.395 & 9.1  & $-0.4$ & 4.6 \\
\end{tabular}
\end{ruledtabular}
\end{table}

We now adopt this hydrogen position in DFT calculations and evaluate exchange couplings using two complementary methods. The first one is superexchange model based on electron hoppings extracted by Wannier fits to the PBE band structure, resulting in~\cite{Mazurenko2006} 
\begin{equation}
 J_i=\frac{4t_{xy\rightarrow xy}^2}{U_{\rm eff}}-\sum_{\alpha}\frac{4t_{xy\rightarrow\alpha}^2J_{\rm eff}}{(U_{\rm eff}+\Delta_{\alpha})(U_{\rm eff}+\Delta_{\alpha}-J_{\rm eff})},
\label{eq:kk}\end{equation}
where $d_{xy}$ is the half-filled orbital of V$^{4+}$, $\alpha$ labels unoccupied $d$-orbitals, and $\Delta_{\alpha}$ is the crystal-field splitting. The first term stands for AFM superexchange arising from electron hoppings between the half-filled orbitals, whereas the second term is FM superexchange due to electron hoppings between the half-filled and empty orbitals. Using the effective Coulomb repulsion of $U_{\rm eff}=4$\,eV and Hund's coupling of $J_{\rm eff}=1$\,eV~\footnote{Note that $U_{\rm eff}$ and $J_{\rm eff}$ apply to V-based
Wannier functions that include both V $3d$ and O $2p$ orbitals, while the DFT+$U$ parameters $U_d$ and $J_d$ mentioned in Sec.~II describe, respectively, the Coulomb repulsion and Hund's coupling in V $3d$ orbitals only.} from Refs.~\cite{Tsirlin2011b,Tsirlin2014}, we arrive at the exchange couplings $J_i^{\rm AFM}$ and $J_i^{\rm FM}$ listed in Table~\ref{tab:exchange}. Additionally, we compute total exchange couplings $J_i$ via the DFT+$U$ mapping approach~\cite{Xiang2011,Tsirlin2014}. 

Both methods arrive at qualitatively similar results. The NN couplings are FM, whereas the NNN couplings are AFM in nature, thus favoring solution-$b$. The largest anisotropy is observed between $J_1'$ and $J_1''$. However, on average these couplings -- both running along $c$ -- differ from $J_1$ (along $b$) by 20\% only. The spatial anisotropy of the NNN couplings is more significant with $J_2/J_2'\simeq 1.8$. The overall energy scale of the NNN couplings, $\bar J_2\simeq 5.6$\,K, is similar to Li$_2$VOSiO$_4$ ($J_2\simeq 5.9$\,K~\cite{Rosner014416}) and Na$_{1.5}$VOPO$_4$F$_{0.5}$ ($\bar J_2\simeq 6.6$\,K) and much lower than in Pb$_2$VO(PO$_4)_2$ with $\bar J_2\simeq 9.3$\,K~\cite{Nath214430} or SrZnVO(PO$_4)_2$ with $\bar J_2=8.6$\,K~\cite{Bossoni014412}. This difference can be traced back to the buckling of the magnetic layers in $AA'$VO(PO$_4)_2$ ($AA'$ = Pb$_2$, SrZn), while the layers are almost or even perfectly flat in NaZnVOPO$_4$(HPO$_4$) (Fig.~\ref{Fig1}), Li$_2$VOSiO$_4$, and Na$_{1.5}$VOPO$_4$F$_{0.5}$.

\section{Discussion}
Our data suggest that NaZnVOPO$_4$(HPO$_4$) is well described by the spin-$\frac12$ FSL model if deformation of the square lattice is taken into account. Although individual data sets such as field-dependent magnetization may be consistent even with a nonfrustrated square lattice, the reduced size of the specific heat maximum, $C_{\rm max}^{\rm mag}/R\simeq 0.41$, indicates the presence of frustration. Indeed, the value of the frustration parameter $f = \frac {|\theta_{\rm CW}|}{T_{\rm N}} \simeq 2.1$ reflects a moderate frustration in the compound. 


In a spin system, the spin-lattice relaxation rate carries information on the low-lying excitations or spin dynamics in the momentum space. 
Typically, $\frac{1}{T_{1}T}$ can be expressed in terms of the dynamic susceptibility $\chi_{M}(\vec{q},\omega_{0})$ as~\cite{Moriya516}
\begin{equation}
	\frac{1}{T_{1}T} = \frac{2\gamma_{N}^{2}k_{B}}{N_{\rm A}^{2}}
	\sum\limits_{\vec{q}}\mid A(\vec{q})\mid
	^{2}\frac{\chi^{''}_{M}(\vec{q},\omega_{0})}{\omega_0},
	\label{t1form}
\end{equation}
where the sum is over the wave vector $\vec{q}$ within the first Brillouin zone, $A(\vec{q})$ is the form-factor of the hyperfine interaction, and $\chi^{''}_{M}(\vec{q},\omega_{0})$ is the imaginary part of the dynamic susceptibility at the nuclear Larmor frequency $\omega _0$. The dominance of different $q$-components [$\vec{q} = 0$ and $\vec{q} = (\pm \pi/a, \pm \pi/b)$] is often visible in the $1/T_1$ data when plotted against temperature, especially for the low-dimensional spin systems with strong exchange coupling. At very high temperatures ($T>J/k_{\rm B}$), $1/T_{1}$ is almost temperature-independent due to uncorrelated moments and can be expressed as~\cite{Nath214430}
\begin{eqnarray}
	\begin{split}
		\lefteqn{\left(\frac{1}{T_{1}}\right)_{T\to \infty} = \frac{(\gamma_{N}g\mu_{\rm B})^{2}\sqrt{2\pi}z^\prime S(S+1)}{3\omega_{\rm ex}}}\\&\qquad&\qquad&\qquad
		\times\frac{\left(\frac{A_x}{z^\prime}\right)^{2}+\left(\frac{A_y}{z^\prime}\right)^{2}+\left(\frac{A_z}{z^\prime} \right)^{2}}{3},
	\end{split}
	\label{t1form_2}
\end{eqnarray}
where $\omega_{\rm ex}=[\max(|J_1|,|J_2|)k_{\rm B}/\hbar]\sqrt{\frac{2zS(S+1)}{3}}$ is the Heisenberg exchange frequency, $z$ is the number of NN spins of each V$^{4+}$ ion, and $z^\prime$ is the number of NN V$^{4+}$ spins of the P(1) site. In the above expression, the hyperfine couplings along different directions are divided by $z^\prime$ in order to account for the coupling of P site with the individual V$^{4+}$ ion. As the measurements are carried out on the powder sample, we have taken the rms average of the couplings along three directions.

For a tentative estimation of the in-plane exchange coupling between the V$^{4+}$ ions we took the high temperature value of $1/T_{11}$ for the strongly coupled P(1) site. Using the experimental parameters obtained for this site ($A^{1}_{x} \simeq 1134$~Oe/$\mu_{\rm B}$, $A^{1}_{y} \simeq -1716$~Oe/$\mu_{\rm B}$, $A^{1}_{z} \simeq 582$~Oe/$\mu_{\rm B}$, $\gamma_N = 108.303\times10^2$~rad.sec$^{-1}$/Oe, $z^{\prime}=4$, $z=4$, $g=2.04$, $S= 1/2$, and $1/T_{11} \simeq 150$ sec$^{-1}$), the magnitude of the maximum exchange coupling strength between V$^{4+}$ ions is estimated to be $\max(|J_1|,|J_2|)\simeq4.2$~K. This value is indeed very close to the dominant in-plane AFM exchange coupling for both the solutions ($a$ and $b$), obtained from the $\chi(T)$ analysis.

It is also instructive to compare the transition temperatures of V$^{4+}$-based FSL magnets. NaZnVOPO$_4$(HPO$_4$) and Li$_2$VOSiO$_4$ feature almost the same $\bar J_2\simeq 6$\,K but different signs of $J_1$ (FM and AFM, respectively), while the former compound has a much lower $T_N\simeq 2.1$\,K than the latter ($T_N\simeq 2.8$\,K~\cite{Melzi1318}). This indicates that FM couplings $J_1$ together with the deformation of the square lattice lead to a visible reduction in $T_N$ in NaZnVOPO$_4$(HPO$_4$). On the other hand, Pb$_2$VO(PO$_4)_2$ shows an even higher $T_N\simeq 3.65$\,K~\cite{Nath214430} because of the 30\% increase in the magnitude of $\bar J_2$. This comparison illustrates that buckling of the [VOPO$_4$] layers caused by variable spacers between the magnetic layers effectively tunes magnetic interactions in FSL-like compounds.


\section{Summary}
We have studied the magnetism of NaZnVOPO$_4$(HPO$_4$) using wide variety of experimental techniques and complementary \textit{ab initio} calculations. With almost flat magnetic layers, this compound shows weaker next-nearest-neighbor couplings ($\bar J_2$) than other V$^{4+}$ phosphates. This reduction in $\bar J_2$ along with the deformation of the spin-$\frac12$ square lattice lead to a lower magnetic transition temperature than in most of the other FSL candidates.


\section {Acknowledgments}
SG and RN would like to acknowledge SERB, India for financial support bearing sanction Grant No.~CRG/2019/000960. SG is supported by the Prime Minister’s Research Fellowship (PMRF) scheme, Government of India. We also acknowledge the support of the HLD at HZDR, member of European Magnetic Field Laboratory (EMFL).
Work at the Ames Laboratory was supported by the U.S. Department of Energy, Office of
Science, Basic Energy Sciences, Materials Sciences and Engineering Division. The Ames Laboratory is operated for the U.S. Department of Energy by Iowa State University under Contract No. DEAC02-07CH11358.


%
					  
\end{document}